\documentclass[superscriptaddress,pra,nofootinbib,twocolumn,notitlepage,10pt]{revtex4-1}
\pdfoutput=1
\usepackage{graphicx}
\usepackage{amsthm}
\usepackage{thmtools}
\usepackage{mathtools}
\usepackage{thm-restate}
\usepackage{amsmath}
\usepackage{amssymb}
\usepackage{comment}
\usepackage{placeins}
\usepackage[caption=false]{subfig}
\usepackage[colorlinks]{hyperref}
\usepackage{tikz}
\usetikzlibrary{calc,backgrounds}

\usepackage{pgffor}
\usepackage{url}
\usepackage{hypcap}
\usepackage{dsfont}
\usepackage{algorithm}
\usepackage{algorithmic}
\usepackage{soul}
\usepackage{diagbox}

\newcommand{\eq}[1]{Eq.~\hyperref[eq:#1]{(\ref*{eq:#1})}}
\renewcommand{\sec}[1]{\hyperref[sec:#1]{Section~\ref*{sec:#1}}}
\newcommand{\app}[1]{\hyperref[app:#1]{Appendix~\ref*{app:#1}}}
\newcommand{\tab}[1]{\hyperref[tab:#1]{Table~\ref*{tab:#1}}}
\newcommand{\fig}[1]{\hyperref[fig:#1]{Figure~\ref*{fig:#1}}}
\newcommand{\figa}[2]{\hyperref[fig:#1]{Figure~\ref*{fig:#1}#2}}
\newcommand{\figx}[2]{\hyperref[fig:#1]{Figure~\ref*{fig:#1}(#2)}}
\newcommand{\thm}[1]{\hyperref[thm:#1]{Theorem~\ref*{thm:#1}}}
\newcommand{\lem}[1]{\hyperref[lem:#1]{Lemma~\ref*{lem:#1}}}
\newcommand{\cor}[1]{\hyperref[cor:#1]{Corollary~\ref*{cor:#1}}}
\newcommand{\defn}[1]{\hyperref[def:#1]{Definition~\ref*{def:#1}}}
\newcommand{\alg}[1]{\hyperref[alg:#1]{Algorithm~\ref*{alg:#1}}}

\def\avg#1{\mathinner{\langle{#1}\rangle}}
\def\bra#1{\mathinner{\langle{#1}|}}
\def\ket#1{\mathinner{|{#1}\rangle}}

\newcommand{\proj}[1]{\ket{#1}\!\!\bra{#1}}

\newcommand{\be}{\begin{equation}}
\newcommand{\ee}{\end{equation}}
\newcommand{\ba}{\begin{eqnarray}}
\newcommand{\ea}{\end{eqnarray}}

\usepackage{color}
\usepackage{xcolor}

\usepackage{colortbl}
\makeatletter

    \def\CT@@do@color{%
      \global\let\CT@do@color\relax
            \@tempdima\wd\z@
            \advance\@tempdima\@tempdimb
            \advance\@tempdima\@tempdimc
    \advance\@tempdimb\tabcolsep
    \advance\@tempdimc\tabcolsep
    \advance\@tempdima2\tabcolsep
            \kern-\@tempdimb
            \leaders\vrule
                    \hskip\@tempdima\@plus  1fill
            \kern-\@tempdimc
            \hskip-\wd\z@ \@plus -1fill }
    \makeatother

\input{Qcircuit}

\begin{document}

\title{Quantum Simulation of Electronic Structure with Linear Depth and Connectivity}

\date{\today}
\author{Ian D.\ Kivlichan}
\affiliation{Google Inc., Venice, CA 90291}
\affiliation{Department of Chemistry and Chemical Biology, Harvard University, Cambridge, MA 02138}
\author{Jarrod McClean}
\affiliation{Google Inc., Venice, CA 90291}
\author{Nathan Wiebe}
\affiliation{Microsoft Research, Redmond, WA 98052}
\author{Craig Gidney}
\affiliation{Google Inc., Santa Barbara, CA 93117}
\author{Al\'{a}n Aspuru-Guzik}
\affiliation{Department of Chemistry and Chemical Biology, Harvard University, Cambridge, MA 02138}
\author{Garnet Kin-Lic Chan}
\email[Corresponding author: ]{gkc1000@gmail.com}
\affiliation{Division of Chemistry and Chemical Engineering, California Institute of Technology, Pasadena, CA 91125}
\author{Ryan Babbush}
\email[Corresponding author: ]{ryanbabbush@gmail.com}
\affiliation{Google Inc., Venice, CA 90291}

\begin{abstract}
As physical implementations of quantum architectures emerge, it is increasingly important to consider the cost of algorithms for practical connectivities between qubits. We show that by using an arrangement of gates that we term the fermionic swap network, we can simulate a Trotter step of the electronic structure Hamiltonian in exactly $N$ depth and with $N^2/2$ two-qubit entangling gates, and prepare arbitrary Slater determinants in at most $N/2$ depth, all assuming only a minimal, linearly connected architecture. We conjecture that no explicit Trotter step of the electronic structure Hamiltonian is possible with fewer entangling gates, even with arbitrary connectivities. These results represent significant practical improvements on the cost of most Trotter based algorithms for both variational and phase estimation based simulation of quantum chemistry.
 \end{abstract}

\maketitle

The electronic structure Hamiltonian describes the properties of interacting electrons in the presence of stationary nuclei. The physics of such systems determine the rates of chemical reactions, molecular structure, as well as the properties of most materials. Toward this goal, multiple approaches to quantum simulating electronic structure have been explored (see e.g. 
\cite{Abrams1997,
Ortiz2001,
aspuru2005simulated,
kassal2008polynomial,
whitfield2010simulation,
Veis2012,
seeley2012bravyi,
Veis2014,
babbush2014adiabatic,
wecker2014gate,
hastings2015improving,
Moll2015,
poulin2015trotter,
babbush2015chemical,
Whitfield2016,
Wilhelm2016,
Sugisaki2016,
McClean2016,
bauer2015,
babbush2016exponentially,
mcclean2016theory,
havlicek2017operator,
reiher2017elucidating,
Motzoi2017,
Bravyi2017,
Rubin2018,
Steudtner2017,
BabbushSparse2}),
with some even demonstrated experimentally \cite{Lanyon2010,
Du2010,
peruzzo2014variational,
Wang2014,
Barends2015,
Shen2015,
Santagati2016,
omalley2016scalable,
Siddiqi2017,
kandala2017hardware,
Dumitrescu2018}.

Most past work has focused on using Gaussian basis functions in second quantization, for which the Hamiltonian contains $O(N^4)$ terms, where $N$ is the number of spin-orbitals. However, a recent paper showed that careful selection of basis functions yields a Hamiltonian with $O(N^2)$ terms \cite{BabbushLow}. While certain bases meeting these conditions are nearly ideal for periodic systems, for single molecules they incur a constant overhead compared to Gaussian bases. In this Letter, we introduce two simulation advances inspired by these recently developed Hamiltonian representations that lower the barrier to practical quantum simulation of chemical systems on emerging hardware
platforms.

Our first result is a new implementation of the Trotter step, which uses an optimal swap network combined with fermionic swap gates in order to avoid the fermionic fast Fourier transform (FFFT), which is costly to implement with restricted qubit connectivity.
Our circuit involves exactly $\binom{N}{2}$ entangling operations for the $\binom{N}{2}$ orbital interactions and is perfectly parallelized to a circuit depth of $N$. We conjecture
that the gate complexity of this Trotter step cannot be improved even with arbitrary connectivity.
Our technique can also be used to simulate Trotter steps of the Hubbard model in $O(\sqrt{N})$ depth, even when restricted to linear qubit connectivity.

Our second result is a new method to prepare arbitrary Slater determinants in gate depth of at most $N/2$ on a linear architecture. This is crucial for preparing initial states in nearly all approaches to quantum simulation, including both variational and phase estimation based algorithms. Our work starts from a known strategy based on the QR decomposition \cite{wecker2015solving}, but organizes the rotations in such a way as to allow the algorithm to run with linear connectivity by using parallelization and elimination of redundant rotations based on symmetry considerations to achieve gate depth of at most $N/2$.

Both algorithms improve asymptotically over all prior implementations specialized to restricted connectivity architectures, and additionally give significant constant factor improvements over the best prior algorithms described with arbitrary connectivity.  Such improvements are crucial when planning simulations with limited hardware resources; thus, we expect these strategies will be useful primitives in near-term experiments. The combination of these two steps enables an extremely low depth implementation of the variational ansatz based on Trotterized adiabatic state preparation \cite{wecker2015progress,BabbushLow} (equivalent to the quantum approximate optimization algorithm when the target Hamiltonian is diagonal \cite{Farhi2014}). Since our Trotter steps appear optimal even for arbitrary connectivities, we expect these results will also prove useful for error-corrected quantum simulations.

\subsection*{Linear Trotter Steps by Fermionic Swap Network}

We consider the general problem of simulating any fermionic Hamiltonian of the form 
\begin{equation}
\label{eq:hamiltonian}
H = \sum_{p q} T_{pq} a^\dagger_{p} a_{q}
+ \sum_p U_p n_p
+ \sum_{p\neq q} V_{pq} n_{p} n_{q},
\end{equation}
where $a^\dagger_p$ and $a_p$ are fermionic creation and annihilation operators and $n_p = a^\dagger_p a_p$ is the number operator.
Mapping to qubits under the Jordan-Wigner transformation \cite{somma2002simulating}, \eq{hamiltonian} becomes (up to constant factors)
\begin{align}
& H = \sum_{p\neq q} \frac{V_{pq}}{4} Z_{p} Z_{q}
- \sum_{p} \left(\frac{T_{pp} + U_{p}}{2} + \sum_{q} \frac{ V_{pq}}{2} \right) Z_{p}\\
& + \sum_{p \neq q} \frac{T_{pq}}{2} \left( X_{p} Z_{p+1} \cdots Z_{q-1} X_{q} + Y_{p} Z_{p+1} \cdots Z_{q-1} Y_{q} \right). \nonumber
\end{align}
This includes a range of Hamiltonians, such as the Hubbard model, finite difference discretization of quantum chemistry, and the dual basis encoding described in \cite{BabbushLow}.

Recent work has shown that single Trotter steps can be implemented for a special case of this Hamiltonian in $O(N)$ depth on a quantum computer with planar nearest-neighbor connectivity~\cite{BabbushLow}. The approach of that work involves (i) applying the FFFT in order to switch between the plane wave basis (where the $a^\dagger_p a_q$ are diagonal single-qubit operators) and the dual basis (where the $n_p n_q$ are diagonal two-qubit operators) and (ii) applying a linear depth swap network which places all qubits adjacent at least once so that the $n_p n_q$ terms can be simulated. That swap network requires $2N$ depth with planar connectivity. We will show a new swap network of $N$ depth with linear connectivity which accomplishes the same result. More importantly, we will show that if one uses fermionic swaps gates instead of qubit swap gates, this swap network will actually enable local simulation of all Hamiltonian terms (the $a^\dagger_p a_q$ terms as well as the $n_p n_q$ terms), still in depth $N$ with linear connectivity. This represents a major improvement over the technique of~\cite{BabbushLow} which requires two costly applications of the fermionic fast Fourier transform per dimension in each Trotter step in order to simulate the $a^\dagger_p a_q$ terms. Additionally, the procedure here is more general since it works for any Hamiltonian of the form of \eq{hamiltonian}.

Fermionic swap gates originated in literature exploring tensor networks for classical simulation of fermionic systems (see e.g. \cite{verstraete2009quantum}). They can be expressed independently of the qubit mapping as
\begin{align}
& f_\text{swap}^{p,q} = 1 + a^\dagger_p a_q + a^\dagger_q a_p - a^\dagger_p a_p - a^\dagger_q a_q\\
& f_\text{swap}^{p,q} a^\dagger_p \left(f_\text{swap}^{p,q}\right)^\dagger = a^\dagger_q
\qquad
f_\text{swap}^{p,q} a_p \left(f_\text{swap}^{p,q}\right)^\dagger = a_q.
\end{align}
Thus, the fermionic swap exchanges orbitals $p$ and $q$ while maintaining proper anti-symmetrization. The importance of exchanging orbitals is related to the qubit representation of the fermionic operators, which under the Jordan-Wigner transformation depends on an ordering of the orbitals called the canonical ordering \cite{Ortiz2001,somma2002simulating}. While interaction terms $n_p n_q$ are 2-local qubit operators under the Jordan-Wigner transform, hopping terms $a^\dagger_p a_q$ are $k$-local qubit operators where $k = |p-q| + 1$. Thus, under the Jordan-Wigner transform, the fermionic swap gate between orbitals $p$ and $p+1$ is a 2-local qubit operator. By applying $|p-q|-1$ such neighboring fermionic swap gates, one can thus bring any two qubits $p$ and $q$ next to each other in the canonical ordering. In our algorithm we will only apply the fermionic swap to neighboring orbitals in the Jordan-Wigner representation; thus, we drop superscripts henceforth and use the notation $f_\text{swap} = \textrm{Jordan-Wigner}[f_\text{swap}^{p,p+1}]$.

The key idea for our algorithm is to construct a nearest neighbor fermionic swap network that is interleaved with gates that simulate the evolution of the Hamiltonian terms within a Trotter-Suzuki decomposition. We construct a fermionic swap network in which each orbital is adjacent in the canonical ordering exactly once. Then, in the layer where orbitals $p$ and $q$ are adjacent in the canonical ordering, 
evolution with the operators $a^\dagger_p a_q + a^\dagger_q a_p$ and $n_p n_q$ can be applied using only 2-local nearest neighbor entangling gates. The entire network can be implemented with exactly  $N$ layers of swaps.

\definecolor{salmon}{HTML}{F8766D}
\definecolor{darkyellowolive}{HTML}{A3A500}
\definecolor{jade}{HTML}{00BF7D}
\definecolor{deepskyblue}{HTML}{00B0F6}
\definecolor{verylightheliotrope}{HTML}{E76BF3}
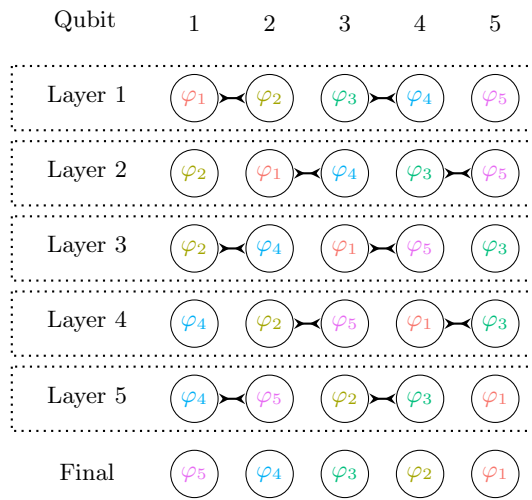
\begin{figure}[ht]
\centering
\begin{tikzpicture}[>=stealth, >-<]
\node [circle,inner sep=0pt, text width=6mm, align=center] (05) at (0,6) {$1$};
\node [circle,inner sep=0pt, text width=6mm, align=center] (15) at (1,6) {$2$};
\node [circle,inner sep=0pt, text width=6mm, align=center] (25) at (2,6) { $3$};
\node [circle,inner sep=0pt, text width=6mm, align=center] (35) at (3,6) {$4$};
\node [circle,inner sep=0pt, text width=6mm, align=center] (45) at (4,6) {$5$};
\node[align=left] at ($(05.north west)+(-1.2,-0.2)$) {Qubit\,};
\node [circle,draw,inner sep=0pt, text width=6mm, align=center] (04) at (0,5) {\color{salmon} ${\varphi_1}$};
\node [circle,draw,inner sep=0pt, text width=6mm, align=center] (14) at (1,5) {\color{darkyellowolive} $\varphi_2$};
\node [circle,draw,inner sep=0pt, text width=6mm, align=center] (24) at (2,5) {\color{jade} $\varphi_3$};
\node [circle,draw,inner sep=0pt, text width=6mm, align=center] (34) at (3,5) {\color{deepskyblue} $\varphi_4$};
\node [circle,draw,inner sep=0pt, text width=6mm, align=center] (44) at (4,5) {\color{verylightheliotrope} $\varphi_5$};
\draw[thick,dotted] ($(04.north west)+(-2.2,0.2)$)  rectangle ($(44.south east)+(0.3,-0.2)$);
\node[align=left] at ($(04.north west)+(-1.2,-0.2)$) {Layer 1};
\node [circle,draw,inner sep=0pt, text width=6mm, align=center] (03) at (0,4) {\color{darkyellowolive}  $\varphi_2$};
\node [circle,draw,inner sep=0pt, text width=6mm, align=center] (13) at (1,4) {\color{salmon} $\varphi_1$};
\node [circle,draw,inner sep=0pt, text width=6mm, align=center] (23) at (2,4) {\color{deepskyblue} $\varphi_4$};
\node [circle,draw,inner sep=0pt, text width=6mm, align=center] (33) at (3,4) {\color{jade} $\varphi_3$};
\node [circle,draw,inner sep=0pt, text width=6mm, align=center] (43) at (4,4) {\color{verylightheliotrope} $\varphi_5$};
\draw[thick,dotted] ($(03.north west)+(-2.2,0.2)$)  rectangle ($(43.south east)+(0.3,-0.2)$);
\node[align=left] at ($(03.north west)+(-1.2,-0.2)$) {Layer 2};
\node [circle,draw,inner sep=0pt, text width=6mm, align=center] (02) at (0,3) {\color{darkyellowolive} $\varphi_2$};
\node [circle,draw,inner sep=0pt, text width=6mm, align=center] (12) at (1,3) {\color{deepskyblue} $\varphi_4$};
\node [circle,draw,inner sep=0pt, text width=6mm, align=center] (22) at (2,3) {\color{salmon} $\varphi_1$};
\node [circle,draw,inner sep=0pt, text width=6mm, align=center] (32) at (3,3) {\color{verylightheliotrope} $\varphi_5$};
\node [circle,draw,inner sep=0pt, text width=6mm, align=center] (42) at (4,3) {\color{jade} $\varphi_3$};
\draw[thick,dotted] ($(02.north west)+(-2.2,0.2)$)  rectangle ($(42.south east)+(0.3,-0.2)$);
\node[align=left] at ($(02.north west)+(-1.2,-0.2)$) {Layer 3};
\node [circle,draw,inner sep=0pt, text width=6mm, align=center] (01) at (0,2) {\color{deepskyblue} $\varphi_4$};
\node [circle,draw,inner sep=0pt, text width=6mm, align=center] (11) at (1,2) {\color{darkyellowolive} $\varphi_2$};
\node [circle,draw,inner sep=0pt, text width=6mm, align=center] (21) at (2,2) {\color{verylightheliotrope} $\varphi_5$};
\node [circle,draw,inner sep=0pt, text width=6mm, align=center] (31) at (3,2) {\color{salmon} $\varphi_1$};
\node [circle,draw,inner sep=0pt, text width=6mm, align=center] (41) at (4,2) {\color{jade} $\varphi_3$};
\draw[thick,dotted] ($(01.north west)+(-2.2,0.2)$)  rectangle ($(41.south east)+(0.3,-0.2)$);
\node[align=left] at ($(01.north west)+(-1.2,-0.2)$) {Layer 4};
\node [circle,draw,inner sep=0pt, text width=6mm, align=center] (00) at (0,1) {\color{deepskyblue} $\varphi_4$};
\node [circle,draw,inner sep=0pt, text width=6mm, align=center] (10) at (1,1) {\color{verylightheliotrope} $\varphi_5$}; 
\node [circle,draw,inner sep=0pt, text width=6mm, align=center] (20) at (2,1) {\color{darkyellowolive} $\varphi_2$};
\node [circle,draw,inner sep=0pt, text width=6mm, align=center] (30) at (3,1) {\color{jade} $\varphi_3$};
\node [circle,draw,inner sep=0pt, text width=6mm, align=center] (40) at (4,1) {\color{salmon} $\varphi_1$};
\draw[thick,dotted] ($(00.north west)+(-2.2,0.2)$)  rectangle ($(40.south east)+(0.3,-0.2)$);
\node[align=left] at ($(00.north west)+(-1.2,-0.2)$) {Layer 5};
\node [circle,draw,inner sep=0pt, text width=6mm, align=center] (0-) at (0,0) {\color{verylightheliotrope} $\varphi_5$};
\node [circle,draw,inner sep=0pt, text width=6mm, align=center] (1-) at (1,0) {\color{deepskyblue} $\varphi_4$};
\node [circle,draw,inner sep=0pt, text width=6mm, align=center] (2-) at (2,0) {\color{jade} $\varphi_3$};
\node [circle,draw,inner sep=0pt, text width=6mm, align=center] (3-) at (3,0) {\color{darkyellowolive} $\varphi_2$};
\node [circle,draw,inner sep=0pt, text width=6mm, align=center] (4-) at (4,0) {\color{salmon} $\varphi_1$};
\node[align=left] at ($(0-.north west)+(-1.2,-0.2)$) {Final};
\draw [thick, >-<] (04) -- (14);
\draw [thick, >-<] (24) -- (34);
\draw [thick, >-<] (02) -- (12);
\draw [thick, >-<] (22) -- (32);
\draw [thick, >-<] (00) -- (10);
\draw [thick, >-<] (20) -- (30);
\draw [thick, >-<] (13) -- (23);
\draw [thick, >-<] (33) -- (43);
\draw [thick, >-<] (11) -- (21);
\draw [thick, >-<] (31) -- (41);
\end{tikzpicture}
\caption{A depiction of how the canonical Jordan-Wigner ordering changes throughout five layers of fermionic swap gates. Each circle represents a qubit in a linear array (qubits do not move) and $\varphi_p$ labels which spin-orbital occupancy is encoded by the qubit during a particular gate layer. The lines in between qubits indicate fermionic swap gates which change the canonical ordering so that the spin-orbitals are represented by different qubits in the subsequent layer. After $N$ layers, the canonical ordering is reversed, and each spin-orbital has been adjacent to all others exactly once.}
\label{fig:swaplayers}
\end{figure}

The swap network is composed of alternating layers of fermionic swaps which reverse the ordering of orbitals as an odd-even transposition sort (parallel bubble sort) run on the reversed list of spin-orbital indices $N$. The first of these two layers consists of fermionic swap gates between the odd-numbered qubits and the even-numbered qubits to their right (qubits $2j+1$ and $2j+2$ for $j \in [0, \lfloor (N-2)/2 \rfloor]$). 
If $N$ is odd, the last qubit is untouched in this layer, because there is no even-numbered qubit to its right. The second of these layers applies a fermionic swap between the even qubits and the odd-numbered qubit to their right (qubits $2j+2$ and $2j+3$, again for $j \in [0, \lfloor (N-2)/2\rfloor]$). In this second layer, the first qubit is always left untouched (there is no even qubit on its left); if $N$ is even the last qubit is untouched. 
Alternating between these layers $N$ times reverses the canonical ordering, thus swapping each spin-orbital with every other spin-orbital exactly once. All layers of this procedure are illustrated for $N=5$ in \fig{swaplayers}.

Suppose that in a particular layer of the swap network, orbital $p$ (encoded by qubit $i_p$) undergoes a fermionic swap with orbital $q$ (encoded by qubit $i_q = i_p+1$). Then, evolution for time $t$ under the fermionic operator $V_{pq} n_p n_q$ and the fermionic operators $T_{pq} (a^\dagger_{p} a_{q} + a^\dagger_{q} a_{p})$ can be performed while simultaneously applying the fermionic swap. This composite two-qubit gate which we refer to as the ``fermionic simulation gate'' can be expressed as
\begin{align}
\label{eq:specialF}
 {\cal F}_t&\left(i_p, i_q \right) = e^{-i V_{pq}  n_{p}n_{q} t} e^{-i T_{pq} (a^\dagger_{p} a_q + a^\dagger_q a_p)t} f_{swap}^{p,q}\\
& = \begin{pmatrix}
1 & 0 & 0 & 0 \\
0 & -i \sin (T_{pq} t) & \cos (T_{pq} t) & 0 \\
0 & \cos (T_{pq} t) & -i \sin (T_{pq} t) & 0 \\
0 & 0 & 0 & -e^{-i V_{pq} t}
\end{pmatrix}\! 
\end{align}
where the second line holds whenever we use the Jordan-Wigner transform and $q = p + 1$. This will always be the case for us due to the reordering of orbitals in our algorithm from the fermionic swap network.

Thus, \fig{swaplayers} depicts an entire first-order (asymmetric) Trotter step if the lines between qubits are interpreted as the gate ${\cal F}_t(i_p, i_q)$. A second-order (symmetric) Trotter step of time $2t$ can be performed by doubling the strength of the final interaction in the first-order step, not performing the corresponding fermionic swap, and then applying all other operations again in reverse order. The network can be extended similarly to higher-order Trotter formulae.  Like any two qubit operation, ${\cal F}_t\left(i_p, i_q \right)$ can be implemented with a sequence of at most three entangling gates from any standard library (e.g. CNOT or CZ) with single-qubit rotations. Finally, the external potential $U_p n_p$ can be simulated by applying single-qubit rotations in a single layer. Interestingly, while charges of the nuclei are all that contribute the external potential (thus, distinguishing various molecules and materials from jellium), these charges enter only through this layer of single-qubit rotations, adding no additional complexity to the quantum circuit for a single Trotter step.

We have shown that exactly $\binom{N}{2}$ two-qubit gates (i.e. fermionic simulation gates ${\cal F}_t(i_p, i_q)$) are sufficient to implement a single Trotter step in gate depth $N$. For $T_{pq}=0$, a Trotter step under \eq{hamiltonian} is equivalent to a network of arbitrary CPhase gates between all pairs of qubits. Since such CPhase networks seem unlikely to simplify, we conjecture that one cannot decompose Trotter steps of \eq{hamiltonian} into fewer than $\binom{N}{2}$ two-qubit gates (assuming no structure in the coefficients). As our gates are fully parallelized, assumption of this conjecture also implies that no algorithm can achieve lower depth for these Trotter steps without additional spatial complexity.

Finally, in \app{hubbard}, we show that the fermionic swap network can be applied to simulate Trotter steps of the Hubbard model on a linear array with $O(\sqrt{N})$ depth. This is an asymptotic improvement in time over all prior approaches to simulate the Hubbard model on a linear array and represents an improvement in space over methods specialized to a planar lattice \cite{verstraete2005mapping}.

\subsection*{Linear Preparation of Slater Determinants with Parallel Givens Rotations}
All schemes for quantum simulation of electronic structure require that one initialize the system register in some state that has reasonable overlap with an eigenstate of interest (e.g. the ground state). Usually, the initial state is a single Slater determinant such as the Hartree-Fock state. This is a trivially preparable computational basis state if the simulation is performed in the basis of Hartree-Fock molecular orbitals. However, as argued in the literature, there is a trade-off between the number of terms in a Hamiltonian representation and the compactness of the Hartree-Fock state \cite{BabbushLow}. Rather than change the basis of the Hamiltonian, which could asymptotically reduce its sparsity, one can use a quantum circuit to rotate the state into the desired basis. Efficient circuits of this kind have previously been considered \cite{Ortiz2001,somma2002simulating}; e.g., \cite{wecker2015solving} describes a procedure for preparing arbitrary Slater determinants with $N^2$ gates using arbitrary connectivity and \cite{BabbushLow} proposes to use the FFFT to prepare a plane-wave state with $O(N)$ depth using planar connectivity. We present here an arbitrary-basis Slater determinant preparation protocol which executes in $N/2$ depth for systems with linear connectivity.

Our scheme is a variant of the QR decomposition based method of constructing single-particle unitaries described in other work \cite{wecker2015solving,Reck1994,Maslov2007}. Any particle-conserving rotation of the single-particle basis can be expressed as
\begin{align}
\label{eq:basis_change}
\widetilde{\varphi}_p & = \sum_{q} \varphi_q u_{pq}
\quad
\tilde{a}^\dagger_p = \sum_{q} a^\dagger_q u_{pq}
\quad
\tilde{a}_p = \sum_{q} a_q u_{pq}^*
\end{align}
where $\widetilde{\varphi}_p$, $\tilde{a}^\dagger_p$, and $\tilde{a}^\dagger_p$ correspond to spin-orbitals and operators in the rotated basis and $u$ is an $N\times N$ unitary matrix. From the Thouless theorem \cite{thouless1960stability}, this single-particle rotation
is equivalent to applying the $2^N \times 2^N$ operator
\begin{align}
  U(u) = \exp\left(\sum_{pq} \left[\log u \right]_{pq} \left(a^\dagger_p a_q - a^\dagger_q a_p\right)\right) \label{eq:unitary}
  \end{align}
where $\left[\log u\right]_{pq}$ is the $(p, q)$ element of the matrix $\log u$. To efficiently implement $U(u)$ without the overhead of Trotterization, we will decompose it into a sequence of exactly $\binom{N}{2}$ rotations of the form
\begin{align}
  R_{pq} \left(\theta\right) = \exp\left[\theta_{pq} \left(a^\dagger_p a_q - a^\dagger_q a_p \right) \right].\label{eq:givens}
\end{align}
In \app{rotation} we show that 
\begin{equation}
\label{eq:correspondence}
R_{pq}\left(\theta\right) U\left(u\right) = U\left(r_{pq}\left(\theta\right) u\right)
\end{equation}
where $r_{pq}(\theta)$ corresponds to a Givens rotation by angle $\theta$ between rows $p$ and $q$ of $u$.

The QR decomposition strategy for decomposing $U(u)$ into a sequence of $R_{pq}(\theta)$ rotations is based on finding a series of $r_{pq}(\theta)$ rotations which diagonalize $u$. This elucidates the inverses of $u$ and $U(u)$ up to some phases:
\begin{align}
\label{eq:diag_strategy}
\left(\prod_k r_k \left(\theta_k\right)\right) u & = \sum_{p=1}^N e^{i \phi_p} \proj{p}\\
\quad
\left(\prod_k R_k \left(\theta_k\right)\right) U\left(u\right) & = \prod_{p=1}^N e^{i \phi_p n_p}
\end{align}
where the index $k$ represents a particular pair of orbitals $p,q$ involved in the rotation at iteration $k$ and $e^{i \phi_p}$ is a unit phase. Given this sequence of rotations and the phases $\phi_p$, we may implement $U$ by applying $\prod_p e^{-i \phi_p n_p}$ (a single layer of gates) and then reversing the sequence of rotations. Viewed in terms of its corresponding action on $u$, \eq{diag_strategy} corresponds to a classical QR decomposition by Givens rotations from \eq{givens}. The right-hand side of \eq{diag_strategy} is the upper-triangular matrix in QR form. But since that matrix is also unitary, the upper-triangular form is diagonal with the $p^\text{th}$ entry equal to $e^{i \phi_p}$.

\begin{figure}
\begin{equation}
\left(\begin{matrix}
* & * & * & * & * & * & * & * & *\\
8 & * & * & * & * & * & * & * & *\\
7 & 9 & * & * & * & * & * & * & *\\
6 & 8 & 10 & * & * & * & * & * & *\\
5 & 7 & 9 & 11 & * & * & * & * & *\\
4 & 6 & 8 & 10 & 12 & * & * & * & *\\
3 & 5 & 7 & 9 & 11 & 13 & * & * & *\\
2 & 4 & 6 & 8 & 10 & 12 & 14 & * & *\\
1 & 3 & 5 & 7 & 9 & 11 & 13 & 15 & *
\end{matrix}\right) \nonumber
\end{equation}
\vspace{-.5cm}
\caption{\label{fig:givens} The numbers above indicate the order in which matrix elements should be eliminated using nearest-neighbor Givens rotations. We see that two elements must be eliminated before any parallelization can begin. Each element is eliminated via rotation with the row directly above it. We place asterisks (*) on the upper-diagonal to emphasize that one only needs to focus on removing the lower-diagonal elements; since the initial matrix and rotations are both unitary, the upper-diagonals will be eliminated simultaneously.}
\end{figure}

When the Givens rotation matrix $r_{pq}(\theta)$ left multiplies the $N\times N$ unitary matrix $u$ it effects a rotation between rows $u_p$ and $u_q$ which can be used to zero out a single element in one of those rows. Since there are $\binom{N}{2}$ elements below the diagonal, the number of Givens rotation required is $\binom{N}{2}$.
The usual strategy for the QR decomposition via Givens rotations involves first rotating all the off-diagonal elements in the first column to zero, and then rotating all the off-diagonal elements in the second column to zero, etc., starting from the bottom. Since Givens rotations affect only the rows that they act upon, one can zero out an entire column before moving on to the next. In order to avoid worrying about non-local Jordan-Wigner strings, we will want to restrict Givens rotations to act on adjacent rows, $q$ and $q-1$. With that restriction, if elements $(p, q)$ and $(p, q-1)$ are already zero, then no Givens rotations between rows $q$ and $q-1$ can restore those elements to nonzero values. This observation suggests a parallelization scheme which is suitable for even a linear array of qubits. The parallelization scheme is illustrated in \fig{givens}.

In the scheme depicted in \fig{givens}, elements should always be eliminated by performing a Givens rotation with the row above it. As we can see from \fig{givens}, one will not perform a Givens rotation to eliminate an element in column $q$ until $2q - 1$ parallel layers of rotations have already occurred. The algorithm terminates once rotations have reached $q = N - 1$; thus, gate depth of $2 N - 3$ is sufficient to implement the basis change.

We can gain additional constant factor efficiencies from symmetries of the Hamiltonian. The usual electronic structure Hamiltonian
has both ${\cal S U}(2)$ (spin) and ${\cal U}(1)$ (particle number $\eta$) symmetry. We can arrange the initial state
to be an eigenstate of spin with the first $N/2$ qubits  spin-up and the remaining qubits spin-down;
$u$ is block-diagonal in these two spin sectors. Performing the procedure in parallel across the two sectors brings the total depth to $N-3$ layers. In addition, working within the $\eta$-electron manifold of Slater determinants, one only needs to perform rotations creating excitations between the $\eta$ occupied orbitals and the $N - \eta$ virtual orbitals. Thus, rather than the $\binom{N}{2}$ Givens rotations required, only $\eta (N - \eta)$ Givens rotations are required. If we assume that the first $\eta / 2$ orbitals of each spin sector are initially occupied, then after $\eta - 1$ parallel steps of the algorithm depicted in \fig{givens}, one has implemented all rotations that couple occupied and virtual spaces (all remaining rotations are between virtual orbitals). If $\eta > N/2$ we can rotate the holes instead of the particles; thus, gate depth of $\eta - 1 < N/2$ is sufficient to prepare any single Slater determinant using our approach.

We have thus shown a method for preparing arbitrary Slater determinants with at most $N/2$ depth on a linear nearest-neighbor architecture. This is even lower depth than any known implementation of the FFFT when the FFFT is restricted to linear or planar connectivities. Thus, our result represents an improvement in situations that call for applying the FFFT on a limited connectivity architecture, such as in the experimental proposal of \cite{BabbushLow}. Unlike implementations of the FFFT based on radix-2 decimation \cite{verstraete2009quantum,BabbushLow}, the state preparation described here is not limited to binary power system sizes.

\subsection*{Conclusion}
We have introduced approaches for both state preparation and time evolution of electronic structure Hamiltonians which execute in at most linear gate depth with linear connectivity. In the near-term, both results raise the prospects
of  practical algorithms for non-trivial system sizes which meet the limitations of available hardware. Even within a fault-tolerant paradigm, both our state preparation and Trotterization procedures afford constant factor improvements over all prior approaches, including those requiring arbitrary connectivity. While we have argued for the optimality of our Trotter steps, proving a formal lower bound remains an open problem. Future work should numerically investigate the Trotter errors associated with these Trotter steps in the spirit of prior work on Gaussian bases \cite{babbush2015chemical,reiher2017elucidating}.

\subsection*{Acknowledgments}
We thank Zhang Jiang, Sergio Boixo, Eddie Farhi, James McClain, Kevin Sung, and Guang Hao Low for helpful discussions. I.\ D.\ K.\ acknowledges partial support from the National Sciences and Engineering Research Council of Canada. A.\ A.-G.\ acknowledges the Army Research Office under Award: W911NF-15-1-0256. We thank contributors to the open source library OpenFermion (\url{www.openfermion.org})~\cite{openfermion} which was used to verify some equations of this work.

\bibliographystyle{apsrev4-1}
\bibliography{science,Mendeley}

\appendix

\section{Hubbard Model Trotter Steps}
\label{app:hubbard}

Using the fermionic swap network described in the main paper, we can also simulate Trotter steps of the 2D Hubbard model with gate depth $O(\sqrt{N})$ on a linear array of qubits. We can do this for Hubbard models with and without spin, but it is currently not clear how one might efficiently handle periodic boundary conditions with the same strategy. Below, we explain how this algorithm would work for the 2D Hubbard model with spins but note that a simple extension of the algorithm is possible for models in $d$ dimensions with gate depth $O(N^{\frac{d-1}{d}})$.
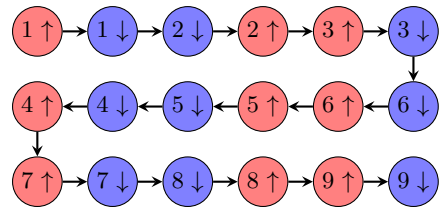
\begin{figure}[ht]
\centering
\begin{tikzpicture}[>=stealth, >-<]
\node [circle,draw,fill=red!50,inner sep=0pt, text width=6mm, align=center] (1u) at (0,0) {$1\uparrow$};
\node [circle,draw,fill=blue!50,inner sep=0pt, text width=6mm, align=center] (1d) at (1,0) {$1\downarrow$};
\node [circle,draw,fill=red!50,inner sep=0pt, text width=6mm, align=center] (2u) at (3,0) {$2\uparrow$};
\node [circle,draw,fill=blue!50,inner sep=0pt, text width=6mm, align=center] (2d) at (2,0) {$2\downarrow$};
\node [circle,draw,fill=red!50,inner sep=0pt, text width=6mm, align=center] (3u) at (4,0) {$3\uparrow$};
\node [circle,draw,fill=blue!50,inner sep=0pt, text width=6mm, align=center] (3d) at (5,0) {$3\downarrow$};
%
\node [circle,draw,fill=red!50,inner sep=0pt, text width=6mm, align=center] (4u) at (0,-1) {$4\uparrow$};
\node [circle,draw,fill=blue!50,inner sep=0pt, text width=6mm, align=center] (4d) at (1,-1) {$4\downarrow$};
\node [circle,draw,fill=red!50,inner sep=0pt, text width=6mm, align=center] (5u) at (3,-1) {$5\uparrow$};
\node [circle,draw,fill=blue!50,inner sep=0pt, text width=6mm, align=center] (5d) at (2,-1) {$5\downarrow$};
\node [circle,draw,fill=red!50,inner sep=0pt, text width=6mm, align=center] (6u) at (4,-1) {$6\uparrow$};
\node [circle,draw,fill=blue!50,inner sep=0pt, text width=6mm, align=center] (6d) at (5,-1) {$6\downarrow$};
\node [circle,draw,fill=red!50,inner sep=0pt, text width=6mm, align=center] (7u) at (0,-2) {$7\uparrow$};
\node [circle,draw,fill=blue!50,inner sep=0pt, text width=6mm, align=center] (7d) at (1,-2) {$7\downarrow$};
\node [circle,draw,fill=red!50,inner sep=0pt, text width=6mm, align=center] (8u) at (3,-2) {$8\uparrow$};
\node [circle,draw,fill=blue!50,inner sep=0pt, text width=6mm, align=center] (8d) at (2,-2) {$8\downarrow$};
\node [circle,draw,fill=red!50,inner sep=0pt, text width=6mm, align=center] (9u) at (4,-2) {$9\uparrow$};
\node [circle,draw,fill=blue!50,inner sep=0pt, text width=6mm, align=center] (9d) at (5,-2) {$9\downarrow$};
%
 \draw [thick, ->] (1u) -- (1d);
 \draw [thick, ->] (1d) -- (2d);
 \draw [thick, ->] (2d) -- (2u);
 \draw [thick, ->] (2u) -- (3u);
 \draw [thick, ->] (3u) -- (3d);
 \draw [thick, ->] (3d) -- (6d);
 \draw [thick, ->] (6d) -- (6u);
 \draw [thick, ->] (6u) -- (5u);
 \draw [thick, ->] (5u) -- (5d);
 \draw [thick, ->] (5d) -- (4d);
 \draw [thick, ->] (4d) -- (4u);
 \draw [thick, ->] (4u) -- (7u);
 \draw [thick, ->] (7u) -- (7d);
 \draw [thick, ->] (7d) -- (8d);
 \draw [thick, ->] (8d) -- (8u);
 \draw [thick, ->] (8u) -- (9u);
 \draw [thick, ->] (9u) -- (9d);
\end{tikzpicture}
\caption{Depiction of the mapping of Hubbard sites to a linear qubit chain. The circles each represent a spin-orbital. As labeled, red circles contain spin-up orbitals and blue circles contain spin-down orbitals. In the Hubbard Hamiltonian, the on-site interaction gives a diagonal couplings between the two spin-orbitals within each spatial orbital (e.g.~$n_{3,\uparrow} n_{3,\downarrow}$) and the hopping terms are off-diagonal between adjacent spatial orbitals of the same spin (e.g.~$a^\dagger_{5,\downarrow} a_{6,\downarrow} + a^\dagger_{6,\downarrow} a_{5,\downarrow}$).
The arrows between the circles indicate the canonical ordering that should be used in the Jordan-Wigner transformation. The general pattern here is that we alternate whether the up or down spin-orbital comes first across the rows, and we alternate whether to order in ascending or descending order across columns.}
\label{fig:hubbard_ordering}
\end{figure}

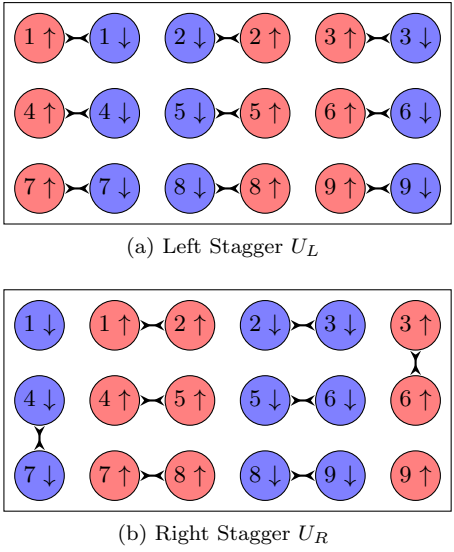
\begin{figure}[h]
\centering
\subfloat[Left Stagger $U_L$]{
\begin{tikzpicture}[>=stealth, >-<,framed]
\node [circle,draw,fill=red!50,inner sep=0pt, text width=6mm, align=center] (1u) at (0,0) {$1\uparrow$};
\node [circle,draw,fill=blue!50,inner sep=0pt, text width=6mm, align=center] (1d) at (1,0) {$1\downarrow$};
\node [circle,draw,fill=red!50,inner sep=0pt, text width=6mm, align=center] (2u) at (3,0) {$2\uparrow$};
\node [circle,draw,fill=blue!50,inner sep=0pt, text width=6mm, align=center] (2d) at (2,0) {$2\downarrow$};
\node [circle,draw,fill=red!50,inner sep=0pt, text width=6mm, align=center] (3u) at (4,0) {$3\uparrow$};
\node [circle,draw,fill=blue!50,inner sep=0pt, text width=6mm, align=center] (3d) at (5,0) {$3\downarrow$};
%
\node [circle,draw,fill=red!50,inner sep=0pt, text width=6mm, align=center] (4u) at (0,-1) {$4\uparrow$};
\node [circle,draw,fill=blue!50,inner sep=0pt, text width=6mm, align=center] (4d) at (1,-1) {$4\downarrow$};
\node [circle,draw,fill=red!50,inner sep=0pt, text width=6mm, align=center] (5u) at (3,-1) {$5\uparrow$};
\node [circle,draw,fill=blue!50,inner sep=0pt, text width=6mm, align=center] (5d) at (2,-1) {$5\downarrow$};
\node [circle,draw,fill=red!50,inner sep=0pt, text width=6mm, align=center] (6u) at (4,-1) {$6\uparrow$};
\node [circle,draw,fill=blue!50,inner sep=0pt, text width=6mm, align=center] (6d) at (5,-1) {$6\downarrow$};
\node [circle,draw,fill=red!50,inner sep=0pt, text width=6mm, align=center] (7u) at (0,-2) {$7\uparrow$};
\node [circle,draw,fill=blue!50,inner sep=0pt, text width=6mm, align=center] (7d) at (1,-2) {$7\downarrow$};
\node [circle,draw,fill=red!50,inner sep=0pt, text width=6mm, align=center] (8u) at (3,-2) {$8\uparrow$};
\node [circle,draw,fill=blue!50,inner sep=0pt, text width=6mm, align=center] (8d) at (2,-2) {$8\downarrow$};
\node [circle,draw,fill=red!50,inner sep=0pt, text width=6mm, align=center] (9u) at (4,-2) {$9\uparrow$};
\node [circle,draw,fill=blue!50,inner sep=0pt, text width=6mm, align=center] (9d) at (5,-2) {$9\downarrow$};
%
 \draw [thick, >-<] (1u) -- (1d);
 \draw [thick, >-<] (2d) -- (2u);
 \draw [thick, >-<] (3u) -- (3d);
 \draw [thick, >-<] (6d) -- (6u);
 \draw [thick, >-<] (5u) -- (5d);
 \draw [thick, >-<] (4d) -- (4u);
 \draw [thick, >-<] (7u) -- (7d);
 \draw [thick, >-<] (8d) -- (8u);
 \draw [thick, >-<] (9u) -- (9d);
\end{tikzpicture}}
\hspace{2cm}
\subfloat[Right Stagger $U_R$]{
\begin{tikzpicture}[>=stealth, >-<,framed]
\node [circle,draw,fill=red!50,inner sep=0pt, text width=6mm, align=center] (1u) at (1,0) {$1\uparrow$};
\node [circle,draw,fill=blue!50,inner sep=0pt, text width=6mm, align=center] (1d) at (0,0) {$1\downarrow$};
\node [circle,draw,fill=red!50,inner sep=0pt, text width=6mm, align=center] (2u) at (2,0) {$2\uparrow$};
\node [circle,draw,fill=blue!50,inner sep=0pt, text width=6mm, align=center] (2d) at (3,0) {$2\downarrow$};
\node [circle,draw,fill=red!50,inner sep=0pt, text width=6mm, align=center] (3u) at (5,0) {$3\uparrow$};
\node [circle,draw,fill=blue!50,inner sep=0pt, text width=6mm, align=center] (3d) at (4,0) {$3\downarrow$};
%
\node [circle,draw,fill=red!50,inner sep=0pt, text width=6mm, align=center] (4u) at (1,-1) {$4\uparrow$};
\node [circle,draw,fill=blue!50,inner sep=0pt, text width=6mm, align=center] (4d) at (0,-1) {$4\downarrow$};
\node [circle,draw,fill=red!50,inner sep=0pt, text width=6mm, align=center] (5u) at (2,-1) {$5\uparrow$};
\node [circle,draw,fill=blue!50,inner sep=0pt, text width=6mm, align=center] (5d) at (3,-1) {$5\downarrow$};
\node [circle,draw,fill=red!50,inner sep=0pt, text width=6mm, align=center] (6u) at (5,-1) {$6\uparrow$};
\node [circle,draw,fill=blue!50,inner sep=0pt, text width=6mm, align=center] (6d) at (4,-1) {$6\downarrow$};
\node [circle,draw,fill=red!50,inner sep=0pt, text width=6mm, align=center] (7u) at (1,-2) {$7\uparrow$};
\node [circle,draw,fill=blue!50,inner sep=0pt, text width=6mm, align=center] (7d) at (0,-2) {$7\downarrow$};
\node [circle,draw,fill=red!50,inner sep=0pt, text width=6mm, align=center] (8u) at (2,-2) {$8\uparrow$};
\node [circle,draw,fill=blue!50,inner sep=0pt, text width=6mm, align=center] (8d) at (3,-2) {$8\downarrow$};
\node [circle,draw,fill=red!50,inner sep=0pt, text width=6mm, align=center] (9u) at (5,-2) {$9\uparrow$};
\node [circle,draw,fill=blue!50,inner sep=0pt, text width=6mm, align=center] (9d) at (4,-2) {$9\downarrow$};
%
  \draw [thick, >-<] (1u) -- (2u);
  \draw [thick, >-<] (2d) -- (3d);
  \draw [thick, >-<] (3u) -- (6u);
  \draw [thick, >-<] (6d) -- (5d);
  \draw [thick, >-<] (5u) -- (4u);
  \draw [thick, >-<] (4d) -- (7d);
  \draw [thick, >-<] (7u) -- (8u);
  \draw [thick, >-<] (8d) -- (9d);
\end{tikzpicture}}
\caption{By repeating the pattern of fermionic swaps shown as $U_L$ and $U_R$ one is able to bring spin-orbitals from adjacent rows next to each other in the canonical ordering so that the hopping term may be applied locally. First, one applies $U_L$. This will enable application of the remaining horizontal hopping term that could not previously be reached. Then, one should repeatedly apply $U_R U_L$. After each application of $U_R U_L$ new vertical hopping terms become available until one has applied $U_R U_L$ a total of $\sqrt{N / 8} - 1$ times. At that point, one needs to reverse the series of swaps until the orbitals are back to their original locations in the canonical ordering. At that point, applying $U_R U_L$ will cause the qubits to circulate in the other direction. This should be repeated for a total of $\sqrt{N / 8} - 1$ times to make sure all neighboring orbitals are adjacent at least once. The total number of layers of fermionic swaps required for the whole procedure is $ \sqrt{9\, N / 2}$.}
\label{fig:hubbard_sorting}
\end{figure}

The 2D Hubbard Hamiltonian with spins is
\begin{equation}
\label{eq:hubbard}
H \! = \! - t \! \sum_{\avg{pq}, \sigma} \left(a^\dagger_{p,\sigma} a_{q,\sigma} + a^\dagger_{q,\sigma} a_{p,\sigma} \right) + U \sum_{p} n_{p,\uparrow} n_{p,\downarrow}
\end{equation}
where $\avg{pq}$ indicates that the sum should be taken over all pairs of spin-orbitals $(p, \sigma)$ and $(q, \sigma)$ which are adjacent on the 2D Hubbard lattice. The Hubbard Hamiltonian is a special case of the general electronic structure Hamiltonian where many of the terms are zero; whereas the general electronic structure Hamiltonian has $O(N^2)$ terms, the Hubbard Hamiltonian has only $O(N)$ terms. The first step in our procedure will be to use the Jordan-Wigner transformation to map \eq{hubbard} to a qubit Hamiltonian. One needs to choose a particular ordering of the orbitals for the Jordan-Wigner transformation in order for our technique to work. The ordering we choose is explained in \fig{hubbard_ordering}.

With the term ordering depicted in \fig{hubbard_ordering}, terms are arranged so that we may immediately simulate all of the $n_p n_q$ terms. The difficult part of this simulation is the hopping terms $a^\dagger_{p,\sigma} a_{q,\sigma} + a^\dagger_{p,\sigma} a_{q,\sigma}$. With the ordering of \fig{hubbard_ordering}, one can also immediately simulate half of the horizontal hopping terms. The final step performs a series of $O(\sqrt{N})$ layers of fermionic swaps depicted in \fig{hubbard_sorting} which cycles all spin-orbitals through configurations in which they are adjacent to all orbitals with which they share a hopping term.

This algorithm would appear to be the most efficient strategy for simulating the 2D Hubbard model on a linear array of qubits. However, note that given planar qubit connectivity, there is an obvious way to implement Trotter steps of $O(1)$ depth that is readily apparent (and likely anticipated by those authors) from the techniques of \cite{verstraete2005mapping}. However, the mapping in Ref.~\cite{verstraete2005mapping} requires doubling the number of qubits in the simulation and involves a more complicated (though still local)
Hamiltonian; this constant overhead may be significant for moderate $N$.

\section{State Preparation by Givens Rotation}
\label{app:rotation}

Here, we provide a pedagogical explanation of the strategy based on Givens rotations discussed in the main text. We will show that one can implement any $2^N \times 2^N$
unitary operator of the form
\begin{align}
  U(u) = \exp\left(\sum_{pq} \left[\log u \right]_{pq} \left(a^\dagger_p a_q - a^\dagger_q a_p\right)\right) \label{eq:thouless}
  \end{align}
where $\left[\log u\right]_{pq}$ is the $(p, q)$ element of the $N \times N$
matrix $\log u$, with a sequence of exactly $\binom{N}{2}$ rotations of the form
\begin{align}
  R_{pq} \left(\theta\right) = \exp\left[\theta_{pq} \left(a^\dagger_p a_q - a^\dagger_q a_p \right) \right].\label{eq:givensR}
\end{align}
Notice that $R_{pq} \left(\theta\right)$ is a special case of the basis transformation unitary $U(u)$ from \eq{thouless} which occurs when $U\left(r_{pq}\left(\theta\right)\right) = R_{pq} \left(\theta\right)$. By the definition of the matrix logarithm we see that
\begin{align}
\label{eq:little_r}
r_{pq}\left(\theta\right) =
\left(\begin{matrix} 
1   & \cdots &    0   & \cdots &    0   & \cdots &    0   \\
\vdots & \ddots & \vdots &        & \vdots &        & \vdots \\
0   & \cdots &    \cos\left(\theta\right)   & \cdots &    -\sin\left(\theta\right)   & \cdots &    0   \\
\vdots &        & \vdots & \ddots & \vdots &        & \vdots \\
0   & \cdots &   \sin\left(\theta\right)   & \cdots &    \cos\left(\theta\right)   & \cdots &    0   \\
                      \vdots &        & \vdots &        & \vdots & \ddots & \vdots \\
0   & \cdots &    0   & \cdots &    0   & \cdots &    1
\end{matrix}\right).
\end{align}
The cosine terms appear in the $p^\text{th}$ and $q^\text{th}$ entries along the diagonal, and the positive (negative) sine term appears at the intersections of row $p$ ($q$) and column $q$ ($p$). We see then that $R_{pq}(\theta)$ represents a $2^N \times 2^N$
matrix whereas $r_{pq}(\theta)$ represents an $N \times N$
matrix. Note that $r_{pq}\left(\theta\right)$ is a Givens rotation matrix.

Crucial to the procedure we will describe is the fact that the map $U(u)$ is a homomorphism under matrix multiplication:
\begin{equation}
\label{eq:homomorphism}
U\left(u_a\right) \cdot U\left(u_b\right) = U\left(u_a \cdot u_b\right).
\end{equation}
We now prove this. To construct our proof, we introduce a representation of a Slater determinant, $C$, which is a matrix whose columns hold the coefficients of the orbitals in some basis $\ket{\phi_i}$. This matrix is an element of the Grassmann algebra.  It is the natural object obtained from a classical mean-field calculation that defines a Slater determinant within the specified basis. The matrix can be computed in two equivalent ways. First, the mapping from this representation to the full space is given by the Pl\"ucker embedding $\Phi$
\begin{align}
\ket{\Phi(C)} = \bigwedge_i \left( \sum_j C^i_j \ket{\phi_j} \right)
\end{align}
where $\wedge$ denotes the Grassmann wedge product.

Second, and more commonly in electronic structure and quantum mechanics, this map can be expressed conveniently in terms of second quantization as
\begin{align}
\ket{\Phi(C)} = \prod_i c_i^\dagger \ket{\emptyset  }
\qquad
c_i^\dagger = \sum_j C^i_j a_j^\dagger
\end{align}
where $\ket{\Phi(C)}$ is in the full Hilbert space, $\ket{\emptyset }$ is the Fermi vacuum, and $a_i^\dagger$ expresses the occupation of an orbital site $\ket{\phi_i}$.  We will first show that the map satisfies
\begin{align}
U(u) \ket{\Phi(C)} = \ket{\Phi(u C)} = \ket{\Phi(\tilde C)}
\end{align}
where we have defined $\tilde C = u C$.  We begin as
\begin{align}
\ket{\Phi(\tilde C)} &= U(u) \prod_i c_i^\dagger \ket{\emptyset }\\ 
& = \prod_i U(u) c_i^\dagger U(u)^\dagger \ket{\emptyset }
= \prod_i \tilde c_i^{\dagger}\ket{\tilde \emptyset } \nonumber
\end{align}
where $\tilde c_i^{\dagger} = U(u) c_i^\dagger U(u)^i$, the rotated vacuum ${\ket{\tilde \emptyset}} = U(u)^\dagger \ket{\emptyset } = \ket{\emptyset }$ due to vanishing action on the vacuum, and we used the fact that anti-Hermitian operators generate the unitary group.   
To demonstrate this equality, we wish to show that
\begin{align}
\tilde c_i^{\dagger} = \sum_{j} \tilde C^{i}_j a_j^\dagger.
\end{align}
Using the BCH expansion to determine $\tilde c_i^{\dagger}$, we find
\begin{align}
\tilde c_i^{ \dagger} &= U(u) c_i^\dagger U(u)^\dagger
= e^{\hat \kappa} c_i^\dagger e^{- \hat \kappa}\\
& = c_i^\dagger + [ \hat \kappa, c_i^\dagger ] + \frac{1}{2} [ \hat \kappa, [\hat \kappa, c_i^\dagger ]] + \cdots \nonumber
\end{align}
where we have defined
\begin{align}
\hat \kappa = \sum_{pq} [\log u]_{pq} a_p^\dagger a_q = \sum_{pq} \kappa_{pq} a_p^\dagger a_q.
\end{align}
Evaluating the first order term, we find that
\begin{align}
[\hat \kappa, c_i^\dagger] & = \left[ \sum_{pq} \kappa_{pq} a_p^\dagger a_q, \sum_j C^i_j a_j^\dagger \right] \\
& = \sum_p \left( \sum_q \kappa_{pq} C^i_q \right) a_p^\dagger; \nonumber
\end{align}
following to higher orders, we find that the effect is to define a new creation operator whose coefficients in the $\ket{\phi_i}$ basis are $e^{\kappa} C^{i}$, i.e.
\begin{align}
\tilde c^{ \dagger}_i = \sum_j \left[ u C^i\right]_j a_j^\dagger,
\end{align}
which demonstrates the equality
\begin{align}
U(u) \ket{\Phi(C)} = \ket{\Phi(\tilde C)} = \ket{\Phi(u C)}.
\end{align}
With this equality, we find that
\begin{align}
U(u_a) U(u_b) \ket{\Phi(C)} & = U(u_a)\ket{\Phi(u_b C)}\\
& = \ket{\Phi(u_a u_b C)}\nonumber
\end{align}
this yields an expansion with coefficients
\begin{align}
\ket{\Phi(u_a u_b C)} & = \prod_i \tilde{\tilde{c_i}}^{\,\dagger} \ket{\emptyset }\\
\tilde{\tilde{c_i}}^{\,\dagger} & = \sum_j \left[u_a u_b C \right]^i_j a_j^\dagger.
\end{align}
From this we see that 
\begin{align}
\ket{\Phi(u_a u_b C)} = U(u_a u_b) \ket{\Phi(C)}
\end{align}
and as the representative $C$ we chose was arbitrary, it must hold for any $C$ within the Grassmann algebra, and thus we conclude that
\begin{align}
U(u_a) U(u_b) = U(u_a u_b)
\end{align}
which shows the desired property.

Combining \eq{little_r} and \eq{homomorphism} brings us to the important observation
\begin{equation}
\label{eq:givensA}
R_{pq}\left(\theta\right) U\left(u\right) = U\left(r_{pq}\left(\theta\right) u\right).
\end{equation}
We will show that by applying a sequence of these rotations, one can implement $U^\dagger$ up to some trivial phases:
\begin{align}
\label{eq:diag_strategyA}
\prod_k R_k \left(\theta_k\right) U\left(u\right) & = \sum_{p=1}^N e^{i \phi_p n_p}\\
\prod_k r_k \left(\theta_k\right) u & = \sum_{p=1}^N e^{i \phi_p} \proj{p}
\end{align}
where the index $k$ represents a particular pair $p,q$ which is applied at iteration $k$ and $e^{i \phi_p}$ is a unit phase. Given this sequence of rotations and the phases defined by $\phi_p$, we may implement $U$ by applying $\prod_p e^{i \phi_p n_p}$ (a single layer of gates) and then reversing the sequence of rotations. We explain how this sequence and these phases can be determined by focusing on how Givens rotations in the smaller space can be used to manipulate $u$. Finding the sequence of rotations in \eq{diag_strategyA} is equivalent to performing the QR decomposition, which involves decomposing a square matrix into a product of an orthogonal (in our case, unitary) matrix right multiplied by an upper-triangular matrix. This upper-triangular matrix is diagonal, with the $p^\text{th}$ entry given by $e^{i \phi_p}$, as in \eq{diag_strategyA}.

When the Givens rotation matrix $r_{pq}(\theta)$ left multiplies the $N\times N$
unitary matrix $u$, the product is a unitary matrix with entries (assuming $p < q$)
\begin{equation}
A_{ij} = \left[r_{pq}\left(\theta\right)u\right]_{ij} =
\begin{cases}
u_{pj} \cos \theta - u_{qj} \sin \theta & i = p\\
u_{qj} \sin \theta + u_{pj} \cos \theta & i = q\\
u_{ij} & \textrm{otherwise}.
\end{cases}\nonumber
\end{equation}
In order to diagonalize $u$ (as shown in \eq{diag_strategyA}) our strategy will always be to use Givens rotations in order to rotate an element $A_{qj}$ to zero. Thus, when applying $r_{pq}(\theta)$ to the matrix $A$, we will always choose $\theta = \arctan(-A_{pj} /A_{qj})$
depending on which column $j$ we are targeting. Because each rotation only modifies two rows of $A$, it is possible to carry out this procedure in parallel to reduce the depth. We discuss an effective strategy for the order of parallel rotations in the main text.

\end{document}